\documentclass[8pt]{article}
\makeatletter
\newcommand{\class@name}{gael}
\makeatother

\usepackage{setspace}
\usepackage{url}      
\usepackage{amsmath}
\usepackage[authoryear,sort&compress]{natbib}
\usepackage[colorlinks,citecolor=blue]{hyperref}
\usepackage{geometry}
\usepackage{titlesec}
\titlelabel{}
\titleformat{\section}{\large\sffamily\bfseries}{}{0pt}{}
\titleformat{\subsection}{\sffamily\bfseries}{}{0pt}{}
\titleformat{\subsubsection}{\sffamily\slshape}{}{0pt}{}

\renewenvironment{abstract}{%
\hfill
\begin{minipage}{0.75\linewidth}
{\sc ABSTRACT}
\smallskip
\hrule\noindent%
}{%
\smallskip
\hrule
\end{minipage}
\bigskip
}

\usepackage{verbatim}
\usepackage{graphicx}
\graphicspath{{pics/}{figs/}}
\def\B#1{\mathbf{#1}}

\begin{document}
\sloppy
\title{\bfseries\sffamily A group model for stable multi-subject ICA on
fMRI datasets}

\author{G. Varoquaux$^{1,3}$%
\footnote{Corresponding author: Gael Varoquaux, 
{\tt gael.varoquaux@normalesup.org},
    Laboratoire de Neuro-Imagerie Assistée par Ordinateur
    NeuroSpin
    CEA Saclay , Bât 145, 91191 Gif-sur-Yvette France
}, 
S. Sadaghiani$^{2,3}$, P. Pinel$^{2,3}$,
A. Kleinschmidt$^{2,3}$, J.B. Poline$^3$, B. Thirion$^{1,3}$
\\
$^1$ Parietal project team, INRIA, Saclay-\^{I}le-de-France, Saclay, France, \\
$^2$ INSERM unit\'e 562, NeuroSpin, Saclay, France \\
$^3$ CEA, DSV, I$^2$BM, Neurospin, Saclay, France
\footnote{Funding from INRIA-INSERM collaboration. 
The fMRI set was acquired in the
context of the SPONTACT ANR project}
}

\date{}
\maketitle 

\pagebreak


\begin{abstract}

Spatial Independent Component Analysis (ICA) is an increasingly-used
data-driven method to analyze functional Magnetic Resonance Imaging
(fMRI) data. 
To date, it has been used to extract sets of mutually correlated brain
regions without prior information on the time course of these regions.
Some of these sets of regions, interpreted as functional
\textit{networks}, have recently been used to provide markers of brain
diseases and open the road to paradigm-free population comparisons.
Such group studies raise the question of modeling subject variability
within ICA: how can the patterns representative of a group be modeled and
estimated via ICA for reliable inter-group comparisons? 

In this paper, we propose a hierarchical model for patterns in
multi-subject fMRI datasets, akin to mixed-effect group models used in
linear-model-based analysis. 
We introduce an estimation procedure, CanICA (Canonical ICA), based on \textit{i)} probabilistic dimension reduction 
of the individual data, \textit{ii)}
%
canonical correlation analysis to identify a data subspace common to
the group \textit{iii)} ICA-based pattern extraction.
In addition, we introduce a procedure based on cross-validation to
quantify the stability of ICA patterns at the level of the group. 
We compare our method with state-of-the-art multi-subject fMRI ICA
methods and show that the features extracted using our procedure are
more reproducible at the group level on two datasets of 12 healthy
controls: a resting-state and a functional localizer study.

\end{abstract}


\section{Introduction}
\label{sec:intro}

Much of our understanding of brain function gained through
functional Magnetic Resonance Imaging (fMRI) has been derived by
correlating experimental signals of task-driven activation with external
stimuli or events. Yet, the covariance structure of the fMRI signal holds
as much information as the paradigm.
Indeed at the scale of functional neuroimaging, positive
correlations between distant regions induced by the coordination of
distributed neuronal activity in a particular experimental context may
provide a fundamental insight into brain function.
Pioneered by \citet{biswal1995}, studies of paradigm-free functional connectivity
investigate correlations in the BOLD (Blood Oxygen Level Dependent)
time series.
Such correlation studies have been decisive in identifying large
functional networks of distant regions\citep{cordes2000,greicius2003}. It has been
established that the functional connectivity signals observed in fMRI
contains neuronal information beyond physiological or scanner noise
\citep{shmuel2008,laufs2003}, so that part of these signals can be considered a trace of
underlying neuronal activity. In addition, recent work suggests that
correlations in the functional signal are shaped by the anatomical
connectivity structure
\citep{honey2007,greicius2008,skudlarski2008,heuvel2009}.

Analysis of the correlation structure of the BOLD signal reveals
large-scale patterns of correlated activity \citep{biswal1995,lowe1998}
and is expected to help understand the subdivision of the brain into
cognitive systems that have coherent activity across time. These systems
may be labeled as \textit{networks} assuming that they result from
underlying brain connections \citep{honey2007,bullmore2009} and serve
distinct functions \citep{fox2007,Smith2009}.

The study of brain function through functional-connectivity mapping
can be carried out without requiring the subject to perform a specific
task. The so-called {\sl resting-state} protocols can be easily
applied, and are especially useful to include impaired subjects in a
multi-group analysis.
The networks identified by such experiments can give insights into
the mechanisms of brain diseases and their modifications in pathological
situation can serve as biomarkers to aid in clinical diagnosis
\citep{greicius2004,wang2006,garrity2007,greicius2008a,mohammadi2009}.
Recently, \citet{seeley2009} have shown that large-scale brain
networks of co-activation are adequate neuro-physiological units to
study the impact of neuro-degenerative diseases.

The study of the correlation structure of brain activity is plagued by
the size of the data: an fMRI dataset comprises more than
$10\,000$ time series associated with the various voxels, which yield
millions of possible pair-wise correlations.
Functional-connectivity analysis has been pioneered through seed-based
studies \citep{biswal1995,cordes2000,fox2005} that potentially uncover
large-scale networks of brain activity correlated with a
user-specified seed region. This approach, although very successful,
is limited by the prior choice of seed
regions of interest.
Various clustering techniques
\citep{cordes2002,thirion2006,golland2007} have been used to
automatically define the regions or signals of interest.
However, spatial ICA \citep{mckeown1998,kiviniemi2003} is, to date, the
most popular method for identifying meaningful patterns in correlation
studies without prior definition of any target region.
The patterns extracted by ICA are usually easy to interpret in a
cognitive neuroscience context, as they are most often well-contrasted,
indicate different underlying physiological, physical, and cognitive
processes, and can often be related to networks observed in different
contexts, such as in seed-based analysis, or cognitive
networks known from the literature \citep{Smith2009}.

ICA extracts salient patterns that are embedded in the data, and are thus
considered as important to describe it.
It is a purely data-driven method based on a loosely-constrained data
model; as a consequence, statistical significance of the extracted patterns
remains unclear. In particular, there is no simple way to extrapolate
from findings obtained in one dataset to other datasets, even if these
are sampled from the same population.
In the context of group analysis, statistically well-controlled
seed-based studies have shown that the BOLD signal contains patterns of
correlation that are highly reproducible across subjects, including in
resting-state experiments \citep{shehzad2009}. Similarly, some patterns
extracted from resting-state fMRI datasets by an exploratory ICA approach
are consistent at the group level \citep{Damoiseaux2006} and have been
used as biomarkers for population comparison
\citep{sorg2007,garrity2007}.

However, ICA patterns can be relatively sensitive to mild data variation.
Various, often non-overlapping, ICA patterns of group-level coherent
activity have been reported, from resting-state data
\citep{Damoiseaux2006,Perlbarg2008a,beckmann2005,kiviniemi2003}, and in task-based
experiments \citep{Calhoun2001a,Beckmann2005a}.
Probabilistic models have been used to provide pattern-level
noise-rejection criteria \citep{Beckmann2004} or goodness-of-fit measures
of the model \citep{Guo2008}, but still they do not provide pattern-level
significance testing. The uncontrolled variability of the individual
patterns is detrimental to population studies: there is no established
framework for between-group comparison or inference on ICA maps.

ICA being an exploratory analysis technique that estimates a mixing
model specific to the data, it is not meaningful to compare directly
patterns estimated on different individual subjects. On the contrary,
group-level patterns can be specialized to each subject
\citep{Calhoun2001a,filippini2009}.  
Different strategies have been adopted for group-level extraction of
ICA patterns. Patterns estimated at the subject level can be
\textit{merged} to form group maps
\citep{Perlbarg2008a, esposito2005} although this is a challenging
task because the correspondence of individual maps may be hard to
assess and the merging operation is difficult to model from a
statistical point of view.
Individual-subject volumes can be concatenated along the time axis to
apply the ICA algorithm on the group data
\citep{Calhoun2001a}. Finally, \cite{Beckmann2005a} have developed a
tensorial extension of ICA that estimates patterns across subjects
sharing the same time course throughout the experiment.

In this paper, we present a novel group model for multivariate patterns
in fMRI volumes and an associated estimation procedure to extract
group-level ICA maps modeling subject variability.  The strength of this
method called CanICA lies in the identification of a subspace of
reproducible components across subjects using generalized canonical
correlation analysis (CCA). Combined with an explicit noise model and
resampling procedure, this enables automatic selection of the number of
components.
In addition, we introduce a cross-validation procedure and metrics to
compare the stability of a set of multi-subject patterns across different
sub-populations.
We compare our method to state-of-the-art fMRI group ICA methods with
different group models: concatenation and tensorial group ICA approaches.
We do not compare to \textit{merging} procedures since they do not rely
on a linear model between individual subject-level datasets and
group-level Independent Components (ICs) and thus cannot be formulated
with a spatially-resolved between-subject variability of group-level ICs.
We show with cross-validation that features extracted by our method are
more stable on a group of 12 controls, both in a resting-state experiment
and in a traditional \textit{activation detection} experiment with a
known paradigm.

\section{Theory}

\subsection{Spatial ICA model for fMRI data}

ICA assumes that the observed data is the linear mixture of
unknown base signals, that are recovered based on measures of statistical
independence. 
The underlying model is that of blind separation of independent sources:
\begin{equation} \B{B} = \B{M}\B{A}, \label{eqn:ICA_model} \end{equation}
where the rows of the matrix $\B{B}$ are the observed patterns, and those
of $\B{A}$ form the patterns corresponding to the estimated independent
sources, and $\B{M}$ is a mixing matrix estimated by ICA. To set the
notations, when considering spatial patterns or components, such as
$\B{A}$ or $\B{B}$, in this article, we will use $n_\text{patterns}
\times n_\text{voxels}$-shaped matrices; $n_\text{patterns}$ is a number
that corresponds to the model order, or possibly the number of
observations, depending on the context.
Note however that the independence of the patterns extracted by ICA
algorithms is not guaranteed in theory and rarely checked in practice,
and that the independence criterion used to identify the components often
boils down to sparse component extraction \citep{Daubechies2009}. There
is no theoretical basis to consider that a pattern is representative of
solely one independent process, for instance movement, although in
practice ICA is so far one of the of factor analytic transformations
\citep{langers2009} most suited to blind pattern extractions from fMRI
data. In addition, it is not clear that, from a neuroscientific point of
view, independence is the right concept to isolate brain networks, as no
functional system is fully segregated.

The patterns $\B{B}$ present in the acquired fMRI volumes are confounded
by observation noise. As a result, the ICA mixing model is most often
applied on a subset of the acquired signal, after an initial
data-reduction step. In most ICA methods, this step is carried out using
a Principal Components Analysis (PCA), the order of which thus determines
the dimension of the signal subspace and thus the number of sources
extracted by ICA. In the context of fMRI data analysis, a probabilistic
PCA model can be used to introduce a noise model as the basis for this
subspace selection \citep{Beckmann2004}. 

\subsection{Existing group models for ICA on multi-subject fMRI data}

ICA is a multivariate analysis technique: voxel-based time courses are not
characterized as such, but as part of signal fluctuations in the
entire brain. As a consequence, the voxel-level group models used in
standard mass-univariate analysis --random effects or mixed effects--
cannot be applied directly ICA patterns. Two main strategies have been
used so far to extract group-level patterns for fMRI images.

A first approach, introduced by \citet{Calhoun2001a}, concatenates
individual subject data and performs data reduction and ICA on the
resulting dataset. Data reduction is done by PCA. Subject maps are
obtained by applying the mixing model learned on the group to the data specific to a subject.
The group model underlying the estimation procedure is that the images
observed $\B{Y}$ in the individual datasets are generated by a mixture
of the group-level ICA patterns $\B{A}$ with additional noise $\B{E}$:
\begin{eqnarray}
    \B{P} & = & \B{M}\B{A}
    \label{eqn:GIFT_level2}
    \\
    \B{Y} & = & \B{W}\B{P} + \B{E}
    \label{eqn:GIFT_level1}
\end{eqnarray}
with $\B{Y} = [\B{Y}_1^T, \dots ,\B{Y}_S^T]^T$ the observed individual
subjects images concatenated along the time axis, and $\B{E}$ the
noise. $\B{W}$ gives a subject-specific set of loadings and $\B{P}$ are
the principle components spanning the inter-subject signal subspace.
This model consists in the addition of a subject-dependent
observation noise in equation \ref{eqn:GIFT_level1} to the ICA model
(equations \ref{eqn:GIFT_level2} or \ref{eqn:ICA_model}). 
The GIFT toolbox \citep{Calhoun2001a} (\url{http://icatb.sourceforge.net/})
and the MELODIC software used in \textit{ConcatICA} mode 
\citep{smith2004} implement variations on this model. 
They differ in the way to represent the group-level patterns: GIFT
uses a T-statistics-based thresholding on a random-effects analysis of
the individually reconstructed maps $\{\B{M}^{-1}
\B{W}_i^{-1}\B{Y}_i,\,i=1\dots S\}$. 
This amounts to building a posteriori a hierarchical model with two
levels of variance. ConcatICA, on the other hand, directly
thresholds the groups patterns (see further).
In either case, the model is estimated in two steps, using principal component
analysis (PCA) to solve equation \ref{eqn:GIFT_level1}, followed
by a noiseless ICA algorithm to identify the independent components in equation
\ref{eqn:GIFT_level2}. In the actual implementations of the estimation
of the group model in equation \ref{eqn:GIFT_level1}, the generative
model differ slightly from the model on a implementation basis,
as data reduction steps may be implemented via successive PCAs to
limit the numerical size of each step.

The tensorial extension to ICA developed by \cite{Beckmann2005a} uses a
model inspired from PARAFAC \citep{harshman1970}: the observed
images $\B{Y}$ are modeled as a trilinear combination of group-level
independent patterns $\B{A}$, common time courses, and subject-specific
loadings, with additional
observation noise $\B{E}$:
\begin{equation}
    \B{Y} = ( \B{M} |\otimes| \B{N} ) \B{A} + \B{E}.
\end{equation}
Matrices $\B{M}$ and $\B{N}$ give the mixing of independent patterns
across subjects and components. $(. |\otimes| .)$ is the Khatri-Rao
product, which imposes a triple-outer-product relationship between
subject components and group-level independent components: subject
components share the same set of independent spatial patterns
and time courses for each spatial map, but the contribution of each
independent pattern varies from subject to subject.
The hypothesis guiding the tensor ICA model is that, in a multi-subject
experiment with external correlates, the time course corresponding to
activation of cognitive networks is set by the experiment, and thus
shared between subjects. On the other hand, tensor ICA may imply only a
low degree of prevalence of the components with respect to the
dataset at hand, because it allows arbitrary subject-level loadings on
each of the independent components.

\citet{Guo2008} have described both approaches as special cases of a 
general decomposition model in which subject-specific images are a linear
combination of group-level independent components:
\begin{equation}
    \B{Y} = \B{M} \B{A} + \B{E}
\end{equation}
with $\B{Y} = [\B{Y}_1^T, \dots \B{Y}_S^T]^T$ the group data matrix made of the
time-wise concatenation of observed individual subjects data. The
Group-ICA model and the Tensor-ICA model can be seen as putting
restrictions on the structure of the mixing matrix $\B{M}$.
\citet{Guo2008} propose a more general estimation algorithm of an
unconstrained mixing matrix with an expectation maximization (EM)
algorithm, to learn the group structure associated via the data. The
limitation of this EM approach is that it is based on modeling the
histogram of the independent components with an mixture of an arbitrary
number of Gaussian components.

\subsection{Group comparison with ICA}

While ICA is often used to compare functional connectivity between
groups, the mixing model of ICA does not provide a natural statistical
framework for comparing patterns estimated from different datasets, unlike
the GLM framework. Two recent contributions have laid out statistical
group-comparison procedures.

\citet{Guo2008} propose a goodness-of-fit measure for their
mixture-of-Gaussian-based ICA generative model using an approximate
likelihood ratio test to compare different mixing models and discriminate if
a group is better modeled as a homogeneous population or as a set of
different subgroups.
However, this approach essentially assesses the amount of data variance
fit by the model (which is measured by the likelihood ratio test) so
that it may be systematically affected by dimension selection issues,
while it assumes that the unmodeled variance is independent across voxels
to allow tractable computations.

\citet{Rombouts2009} use Tensor-ICA and a two sample t-test on the
loadings of the different group patterns between healthy controls and
patients with dementia to identify the patterns represented unevenly in
the two populations.

The above procedures provide important indications on data structure, but
they do not directly highlight the difference between groups in the
individual ICs. Some of these individual patterns represent
biologically-meaningful components and are used in cognitive studies and
as biomarkers. There is thus a strong interest in basing group
comparisons on the patterns themselves, rather than the complete mixing
model, such as performed in \cite{garrity2007,greicius2004,mohammadi2009}. 
The difficulty in comparing these patterns stems from the fact that
the estimation performed by the ICA algorithm is not robust against mild
data variation: some global differences may exist between two
decompositions of datasets that resemble each other; for instance, an
IC present in one dataset can show salient features that are separated
into two components in another dataset, so that we may consider the IC as
\textit{split} into two in the other dataset.

\section{Materials and methods}

\subsection{A multivariate extension of mixed-effects models}

\label{sec:model}

To achieve a better control on variability of the ICs due to individual
difference between subjects, we introduce a generative model of the
signal making the separation between subject-to-subject variability, and
observation noise specific to a subject or a session. At the group level,
we describe the BOLD signal by a set of patterns $\B{A}$ corresponding to
different independent components. These different components are
extracted in a signal space common to the group and spanned by principal
components $\B{B}$. The generative model relates these group-level
patterns to the observed signal via different noise terms.

\subsubsection{Generative model}

\paragraph{Group-variability model}

\begin{figure*}
  \begin{center}
   \includegraphics[width=\linewidth]{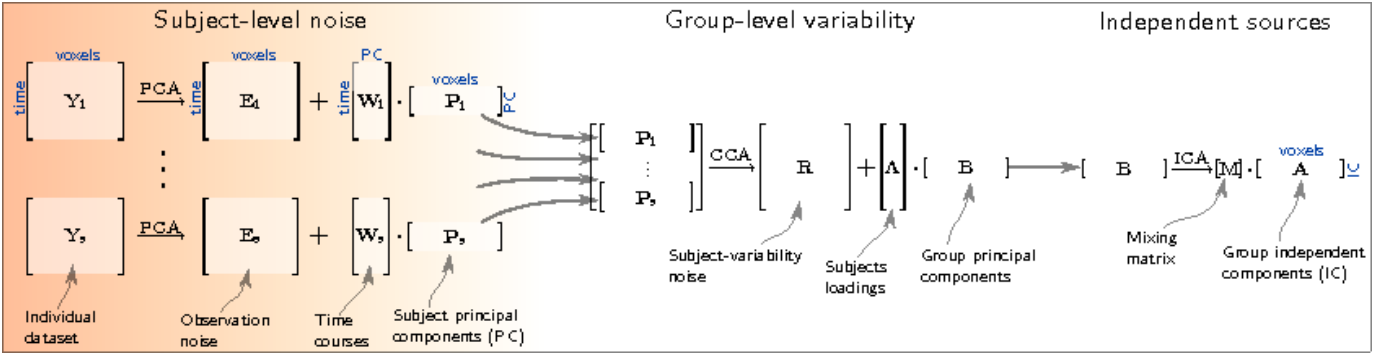}
   \caption{Diagram of the estimation
	    procedure: successive estimation steps are applied from left to right,
	    starting from individual datasets, to end with group-level
	    independent components.
	\label{fig:diagram_pretty}
   }
  \end{center}
\end{figure*}

The activity recorded on each subject $s$ can be described by a set of
subject-specific spatial patterns $\B{P}_s$, which are a combination of
the group-level patterns $\B{B}$ and additional subject-variability:
\begin{equation}
    \text{for each subject $s$,} \quad
    \B{P}_s = \B{\Lambda}_s \, \B{B} + \B{R}_s,
\end{equation}
with $\B{\Lambda}_s$ a loading matrix giving how much each pattern is
represented in subject $s$, and $\B{R}_s$ a residual matrix giving the
deviation from the group patterns. In other words, a group
description can be written considering the group of patterns 
(vertically concatenated matrices) $\B{P} = \{\B{P}_s\}$, 
$\B{R} = \{\B{R}_s\}$, and $\B{\Lambda} = \{\B{\Lambda}_s\}$, $s = 1 \dots S$,
\begin{equation}
    \B{P} = \B{\Lambda} \, \B{B} + \B{R}.
\label{eq:subject_variability}
\end{equation}
$\B{\Lambda}$ is a loading matrix relating the subject-level components
for each subject to the group-level components. If there are
$n_\text{grp}$ components generating the signal at the group level, and
each of the $S$ individual datasets are described by 
$n_\text{sbj}$, the shape of the loading matrix $\B{\Lambda}$
is $(S \cdot n_\text{sbj}, n_\text{grp})$ (see figure 
\ref{fig:diagram_pretty} for a diagram).

\paragraph{Observation model}

For each acquisition-frame time point the observed data is a combination
of different subject-specific patterns $\B{P}_s$ confounded by
observation noise: let $\B{Y}_s$ be the resulting spatial images in BOLD
MRI sequences for subject $s$ (an $n_\text{frames} \times
n_\text{voxels}$ matrix), $\B{E}_s$ the observation noise, and $\B{W}_s$
a loading matrix such that:
\begin{equation}
    \B{Y}_s =  \B{W}_s \, \B{P}_s + \B{E}_s.
\label{eq:observation_noise}
\end{equation}

\subsubsection{Parallel to mixed-effect models}

The generative model can be written in a non-hierarchical form as:
\begin{equation}
\left[
\begin{matrix}
    \mathbf{Y}_1
    \\
    \vdots
    \\
    \mathbf{Y}_s
\end{matrix}
\right]
= 
\left[
\begin{matrix}
\mathbf{W}_1 & 0 & 0 \\
0 & \ddots & 0 \\
0 & 0 & \mathbf{W}_s 
\end{matrix}
\right]
\left(
\left[
\begin{matrix}
    \mathbf{\Lambda}_1
    \\
    \vdots
    \\
    \mathbf{\Lambda}_s
\end{matrix}
\right]
\mathbf{M} 
\mathbf{A} 
+
\left[
\begin{matrix}
    \mathbf{R}_1
    \\
    \vdots
    \\
    \mathbf{R}_s
\end{matrix}
\right]
\right)
\quad
+
\left[
\begin{matrix}
    \mathbf{E}_1
    \\
    \vdots
    \\
    \mathbf{E}_s
\end{matrix}
\right]
\end{equation}
In the group ICA framework of \citet{Guo2008},
\begin{align}
\B{Y} &= \tilde{\B{M}} \B{A} + \tilde{\B{E}}
\label{eqn:guo}
\\\text{with}\quad
\tilde{\B{M}} &= 
\left[
\begin{matrix}
\mathbf{W}_1 & 0 & 0 \\
0 & \ddots & 0 \\
0 & 0 & \mathbf{W}_s 
\end{matrix}
\right]
\left[
\begin{matrix}
    \mathbf{\Lambda}_1
    \\
    \vdots
    \\
    \mathbf{\Lambda}_s
\end{matrix}
\right]
\mathbf{M}
\label{eqn:m_tilde}
\\\text{and}\quad
\tilde{\B{E}} &=
\left[
\begin{matrix}
\mathbf{W}_1 & 0 & 0 \\
0 & \ddots & 0 \\
0 & 0 & \mathbf{W}_s 
\end{matrix}
\right]
\left[
\begin{matrix}
    \mathbf{R}_1
    \\
    \vdots
    \\
    \mathbf{R}_s
\end{matrix}
\right]
+
\left[
\begin{matrix}
    \mathbf{E}_1
    \\
    \vdots
    \\
    \mathbf{E}_s
\end{matrix}
\right]
\label{eqn:e_tilde}
\end{align}
In other words, the mixing matrix can be factored out in a subject-specific
matrix and group-level matrix, and the rejected noise is the addition of
two terms, a subject-specific one and a group-level one, that explain
different components of the signal. 
If we compare the ICA analysis with a GLM analysis, as suggested in
\cite{mckeown1998}, in equation \ref{eqn:guo}, $\tilde{\B{M}}$
corresponds to the design matrix, and $\tilde{\B{E}}$ to the residuals.
We can see that the decomposition of these matrices in equations
\ref{eqn:m_tilde} and \ref{eqn:e_tilde} can also be identified as the 
non-hierarchical expression of corresponding terms in a mixed-effect
model (see for instance \citet{Friston2005}).

Thus, the two-level group model exposed above can be seen as a
multivariate formulation of mixed-effects models, often used in a
standard univariate GLM-based analysis, as it models two sources of
variance.  However, it is not hypothesis-driven and does not rely on
the knowledge of external correlates.

\subsection{Estimation procedure} 

In this section, we describe the estimation procedure used to extract the
patterns of interest from the data. The different variables estimated
corresponding to the generative model are noted with a hat: if $\B{A}$ is
the population set of patterns introduced in the model, $\hat{\B{A}}$
denotes the corresponding patterns estimated from the data. Figure 
\ref{fig:diagram_pretty} gives a diagram summarizing the model and the
successive estimation steps.

\subsubsection{Noise rejection using the generative model}

We start from fMRI image sequences, for each subject, corrected for
delay in slice acquisition and for motion, then registered to a common
template space.  We extract the time series corresponding to a mask of
the brain, center and variance-normalize them, resulting in pattern
matrices $\{\B{Y}_s, s=1...S\}$, with $s$ corresponding to the subject
index. We separate reproducible group-level patterns from noise by
estimating successively each step of the above hierarchical model.

\paragraph{Subject-level identification of observation noise}

First, we separate observation noise $\B{E}_s$ from subject-specific
patterns $\B{P}_s$ as in equation (\ref{eq:observation_noise}),
through principal component analysis (PCA). The principal components
explaining most of the variance for a given subject's data set form
the patterns of interest, while the tail of the spectrum is considered
as observation noise.

Specifically, for each subject, we use a singular value decomposition
(SVD) of the individual data matrix, 
\begin{equation}
    \B{Y}_s = \B{U}_s \, \B{\Sigma}_s \, \B{V}_s.
    \label{subject_SVD}
\end{equation}
We retain the first $n_\text{sbj}(s)$ rows of $\B{V}_s$ to 
constitute the patterns $\B{\hat{P}}_s$ and the first $n_\text{sbj}(s)$ columns 
of $(\B{U}_s \, \B{\Sigma}_s)$ give their loadings $\B{\hat{W}}_s$ in the 
subject's data, that can be interpreted as time courses of the components.
The residual constitutes the observation noise, $\B{\hat{E}}_s$: 
\begin{eqnarray}
    \B{\hat{P}}_s & = & (\B{V}_s)_{1\dots n_\text{sbj}}, \\
\label{eq:indiv_patterns}
    \B{\hat{W}}_s & = & (\B{U}_s \, \B{\Sigma}_s)_{1\dots n_\text{sbj}}, \\
\label{eq:indiv_loadings}
    \B{\hat{E}}_s & = & 
	\B{Y}_s - \B{\hat{W}}_s \, \hat{\B{P}}_s,
\end{eqnarray}
where $n_\text{sbj}(s)$ is the number of extracted components describing the
signal at the subject level for the subject $s$.

Setting the number $n_\text{sbj}(s)$ of principal components to retain at the subject level
is a difficult question and not the main focus of this article. In
general, this number can be selected by analyzing the individual
datasets. We have
found that with information-theoretic criteria initially used for
model-order selection in PCA algorithms (\citet{Minka2001,
Beckmann2004,Calhoun2001a}), the number of identified sources is
overestimated for long fMRI time series and many non-meaningful ICA
components are extracted. Various studies have been conducted of the
influence of PCA-model-order choice in fMRI ICA methods
\citep{himberg2004,li2007,kiviniemi2009}. The optimal number of ICs has
been empirically found to lie between 20 and 60 components depending
on the dataset.  

We use the resampling method described by
\cite{mei2008} in the context of statistical shape modeling. 
This method estimates the number of principal components necessary to
model the data in the presence of Gaussian-distributed noise by
assessing the stability of the subspaces spanned by the first
$n_\text{sbj}(s)$
principal components. 
\cite{mei2008} showed empirically that it does not lead to diverging model order
when new datasets are generated from the selected principle components
with additional Gaussian-distributed noise. For more details on the
model-order selection algorithm and its robustness to noise, we refer the
reader to \cite{mei2008}.
We apply this algorithm on the individual subject
datasets, $\B{Y}_s$ after the various prepocessing steps, 
as it is input in our group
model. For computational reasons, we estimate the subject-level 
number of components on only one subject's data if all the individual
datasets share the same acquisition parameters and length and use this
parameter for all individual datasets: we assume that $n_\text{sbj}(s)$
is independent of $s$, and we write $n_\text{sbj}$.

\paragraph{Canonical correlation analysis to estimate the group-level
patterns}

At the group level, we are interested in identifying the sub-space
common to each subject's patterns. For this purpose, we use
generalized Canonical Correlation Analysis (CCA).

CCA is most often used to compare two multivariate datasets by finding
the successive univariate components pairs of each dataset that maximize
cross-correlation. The components pairs are called {\sl canonical
variables}, and the associated correlation is the {\sl canonical
correlation}. We use a generalization of CCA to multiple datasets
\citep{Kettenring1971,krzanowski1979}-- in our case, one dataset per subject. 
We start from the different whitened datasets%
\footnote{The singular
value spectrum of $\hat{\B{P}}_s$ is made only of ones:
$\hat{\B{P}}_s \hat{\B{P}}_s^T =
\B{1}$. In other words, the amount of variance each IC wich
accounts for at the subject level is not modeled at the group level:
only the patterns are retained, $\hat{\B{\Sigma}}_s$ is not considered.}
$\hat{\B{P}}_s$, and concatenate them to form $\hat{\B{P}}$. We
perform an SVD on $\hat{\B{P}}$:
\begin{equation}
    \B{\hat{P}} = \B{\Upsilon} \, \B{Z} \, \B{\Theta}.
\end{equation}
where $\B{\Upsilon}$ and $\B{\Theta}$ are rotation matrices, and $\B{Z}$
is the diagonal matrix of singular values. The rows of 
$\B{\Theta}$ form the canonical variables, in other words the
inter-subject reproducible components, and the singular values 
on the diagonal of the matrix $\B{Z}$ form the canonical
correlations, which yield a measure of between-subject reproducibility.
We retain the first $n_\text{grp}$ vectors of $\B{\Upsilon}$, the 
canonical weights forming the patterns of interest at the group level:
\begin{equation}
    \B{\hat{B}}_s = (\B{\Upsilon}_s)_{1\dots n_\text{grp}},
\end{equation}
We select the dimension $n_\text{grp}$ by keeping only the canonical variables
for which the corresponding canonical correlation $\B{Z}$ is above a
significance threshold as described below.

The significance threshold on the canonical correlations is set by
sampling a bootstrap distribution of the maximum canonical correlation
using $\B{\hat{E}}_s$, the subject-level observation noise identified
previously, instead of the subject-level components of interest
$\B{\hat{P}}_s$. 
Selected canonical variables have a probability $p < 0.05$ of being
generated by data consisting only of observation noise. This
procedure corresponds thus to keeping only the group-level components
more reproducible than observation noise. We give more details on this
procedure in the supplementary materials.

\bigskip

The estimation procedure for the group-level subspace of interest is
thus done by minimizing the amount of unexplained signal in each
subject while using a fixed number of components, and then by
maximizing the subspace stability at the group level. 
Both optimizations are performed using SVD. For each step,
the unexplained variance is chosen based on a noise model.

\subsubsection{Group-level independent components}

The above estimation procedure selects group-level components $\B{B}$
spanning the subspace of common patterns of activation. We apply FASTICA
\citep{Hyvarinen2000} on this subspace, to separate group-level
independent sources $\B{A}$, estimating the mixing model
\ref{eqn:ICA_model}.

Finally, as the interesting features of the ICA patterns lie in blobs
standing out from the background, it is important to threshold the
resulting map to keep only the tail of the intensity
distribution\footnote{Although this is sometimes done implicitly by
simply referring to the brain regions that stand out when looking at
spatial maps, it is important to this do this reduction explicitly.}. 
For this, we use as simple null-hypothesis distribution a normal
distribution of unit standard deviation. Indeed, the FastICA algorithm
works on whitened data (thus of isotropic unit variance) and
estimates maximally non-Gaussian directions. As in \cite{Schwartzman09},
this null hypothesis models the central mode of the maps, and we select
the voxels for which the absolute intensity exceeds a fixed threshold,
that corresponds to a certain level of specificity with respect to null
distribution. More complex thresholding approaches have been developed,
for instance based on mixture modeling \citep{Beckmann2004}. Our method
leads to fewer selected voxels on a pattern with very few salient
features, such as on artifact patterns. In addition, it is consistent
with the FastICA model.

\bigskip

The main differences of our estimation method compared to the most
currently-used fMRI ICA methods (GIFT and MELODIC) are:
\begin{enumerate}
    \item Model-order determination by rejecting components that can be
    generated by Gaussian noise.
    \item Selecting a subspace of reproducible components at the group
    level using canonical correlation analysis as a device to separate
    subject-level observation noise for group variability.
    \item Thresholding ICs based on the absolute value of voxel
    intensity.
\end{enumerate}
In this article, we are mostly interested in discussing the importance of
point 2, which is the expression of the group model used to describe
between-subject variability.

\subsection{Cross-validation of group-level patterns}

The use of ICA is motivated by the fact that the patterns extracted
from the fMRI data display meaningful features in relation to our
knowledge of functional neuroanatomy. The validation criteria for an
ICA decomposition are unclear, as this algorithm is not based on a
testable hypothesis.  However, for the method to be usable in group
studies, the features extracted from a group of healthy controls
should be comparable between different subgroups of subjects, and
generalize to different subgroups.

To test the reproducibility of the results across groups of subjects,
we split our group of subjects in two and learn ICA maps from each
sub-group: this yields the sets of patterns $\B{A}_1$ and
$\B{A}_2$. We compare the overlap of thresholded maps and reorder one
set to match maps by maximum overlap. Reproducibility can be
quantified by studying the cross-correlation matrix $\B{C} = \B{A}^T_1
\B{A}_2$. For unit-normed components, $\B{C}_{i,j} = 1$ if and only if
$(\B{A}_1)_i$ and $(\B{A}_2)_j$ are identical.

To quantify the reliability of the patterns identified on the full
datasets, we select for each pattern extracted from the full dataset the
best matching one in the different subsets computed in the
cross-validation procedure using Pearson's correlation coefficient. Along
with the extracted maps, we report the average value of this
pattern-reproducibility measure.

\paragraph{Validation experiments: comparing to other methods}

We compare our method with state-of-the-art methods using different
group-model estimation procedures: concatenation approaches using the
implementations of the GIFT ICA toolbox \citep{Calhoun2001a} and the
MELODIC software \citep{smith2004}, as well as the tensor ICA model, also
implemented in the MELODIC software \citep{Beckmann2005a}.

Also, to isolate the effect of separating subject-level variance from
group-level variance in the 
estimation algorithm from implementation-specific details, we run a modified
version of CanICA without the CCA step, 
retaining the subject-level variance $\B{\Sigma}_s$ at
the group level on the same datasets: equations \ref{eq:indiv_patterns} 
and \ref{eq:indiv_loadings} are thus replaced by
  \begin{eqnarray}
    \B{\tilde{P}}_s & = & (\B{\Sigma}_s \B{V}_s)_{1\dots n_\text{sbj}}, \\
    \B{\tilde{W}}_s & = & (\B{U}_s)_{1\dots n_\text{sbj}}.
\label{eq:indiv_pattern_nocca}
\end{eqnarray}
With this estimation algorithm, the variance explained by the components
at the subject level is carried on to the group level. This is analogous
to a fixed-effect model.

As estimating larger numbers of independent components tend to explore
less stable patterns, we run all analysis using the number of
components estimated by CanICA.
We perform cross-correlation analysis on the non-thresholded ICA maps, but
also use each implementation's thresholding algorithm to separate the
features of interest. In an effort to separate the impact of the group
model from the impact of the thresholding heuristic, we study
reproducibility of non-thresholded maps, as well as thresholded maps,
although non-thresholded maps can be considered as of little
neuro-scientific interest. The mixture model implemented in MELODIC
selects 3000 voxels on average on all the patterns. We set the
specificity for the other methods to select the same average number of
voxels. For GIFT, the thresholding is done on the $t$ statistics
maps generated by the algorithm. We threshold at $|t|>2$.

\paragraph{Similarity measures for decompositions learned in different subjects}

We define two measures to quantify the stability of the subspace
spanned by the components and the reproducibility of the maps. First,
a measure of the overlap of the subspaces selected in both groups is
given by the energy of the matrix, i.e. its Frobenius norm: $E =
\text{tr}\,({\B{C}^T
\B{C}})$. To compare this quantity for different subspace sizes, we
normalize it by the minimum dimension of the subspaces,
\begin{eqnarray}
d & = & \min(\text{rank}\,\B{A}_1, \text{rank}\,\B{A}_2) \\
e & = & \frac{1}{d}
\text{tr}\,({\B{C}^T \B{C}}).
\end{eqnarray}
$e$ quantifies the reproducibility of the subspace spanned by the maps.
For $e=1$, the two groups of  maps span the same subspace, although
individual independent components may differ. If $\B{A}_1$ and $\B{A}_2$
have different dimensions, taking the minimum dimension of both does not
account for possible instability of patterns extracted from a group and
not the other. However, the dimensionality as estimated by our procedure
does not vary much over the cross-validation pairs (maximum 10\% relative
difference).

Second, we measure one-to-one reproducibility of maps. 
We reorder the matrix $\B{C}$ by matching sequentially maximally mutually
correlated components of $\B{A}_1$ and $\B{A}_2$, considering the
absolute value of the pair-wise correlation. This procedure creates
a reordered cross-correlation matrix $\B{\tilde{C}}$ with maximal
matching values on the diagonal. If one of the two groups 
has more components than the other, we use the smallest of the two as a
reference, and stop the matching procedure once all of its components
have been matching. As a result, $\B{\tilde{C}}$ is a square matrix, that
we populate with absolute values of pair-wise correlations.
Although we acknowledge that this algorithm may not give the optimal
matching in general,
we observe that the solution is satisfactory in practice.
We use the normalized trace of the reordered cross-correlation matrix, 
\begin{equation}
t =\frac{1}{d}\text{tr}\,(\B{\tilde{C}}),
\end{equation}
as an overall measure of one-to-one overlap between matched pairs of
components.

\subsection{fMRI datasets}
\label{sec:material}

We apply the probabilistic ICA method described above to extract
brain networks from two multi-subjects datasets: a group
resting-state study and a functional localizer. 

\paragraph{Resting-state data}
We use datasets from a resting-state experiment: subjects were
blindfolded and instructed to keep their eyes closed. The resting state
session was recorded as the first session preceding a series of cognitive
experiments as part of a study not detailed here. At the time of
resting-state data collection, subjects were naive with respect to the
nature of the subsequent experiments. 
Twelve right-handed healthy volunteers (2 female; ages 19--30) gave
written informed consent before participation in imaging on a 3T MRI
whole-body scanner (Tim-Trio, Siemens, Erlangen). The study received
local ethics committee approval. 820 EPI volumes (25 slices, $TR =
1.5\,\text{s}$, $TE = 30\,\text{ms}$, FOV $19.2\,\text{cm}\times
19.2\,\text{cm}$, $3\,\text{mm}$ isotropic resolution) were acquired
during a rest period of 20 minutes. A full description of the paradigm
and the acquisition parameters can be found in \citet{sadaghiani2009}. 

\paragraph{Functional localizer data}
We used an event-related experimental paradigm consisting of ten
conditions. Subjects were presented with a series of stimuli and were
engaged in tasks such as passive viewing of horizontal or vertical
checkerboards, left or right button press after auditory or visual
instruction, computation (subtraction) after video or visual instruction
and sentence processing, from the auditory or visual modality. Events
were randomly occurring in time (mean inter stimulus interval:
$3\,\text{s}$), with ten occurrences per event type (except button
presses for which there are only five trials per session). 

Two hundred right-handed subjects participated in the study. The subjects
gave informed consent and the protocol was approved by the local ethics
committee. Functional images were acquired on a 3T Bruker scanner using
an EPI sequence (40 slices, $ TR = 2.4\,\text{s}$, $TE = 60\,\text{ms}$,
FOV $19.2\,\text{cm} \times 19.2\,\text{cm}$, $3\,\text{mm}$ isotropic
resolution). A session comprised $132$ scans. The first four functional
scans were discarded to allow the MR signal to reach steady state. The
experimental paradigm and the acquisition parameters are described with
more detail in \cite{Pinel07}.
To balance the comparison with resting-state data, we first report
reproducibility analysis on $12$ subjects acquired consecutively. We also
study reproducibility on a group of $62$ subjects all acquired
successively in time.

\paragraph{Preprocessing}

The above datasets were preprocessed using the SPM5 software (Wellcome
Department of Cognitive Neurology; http://www.fil.ion.ucl.ac.uk/spm).
After slice-timing interpolation and motion correction, cerebral volumes
were realigned to the MNI152 inter-subject template and smoothed with a
$5\,\text{mm}$ isotropic Gaussian kernel. Voxels contained in a mask of
the brain (approximately $40\,000$ voxels) were selected for pattern
extraction using the CanICA procedure. We do not apply a frequency filter
to the time-series, as we have found out that this yields ICs displaying
better separation between brain networks and artifacts such as movement
or blood-flow related patterns.

\section{Results}
\label{sec:results}

\subsection{Group-level patterns extracted}

\begin{figure*}
  \begin{center}
   \includegraphics[width=\linewidth]{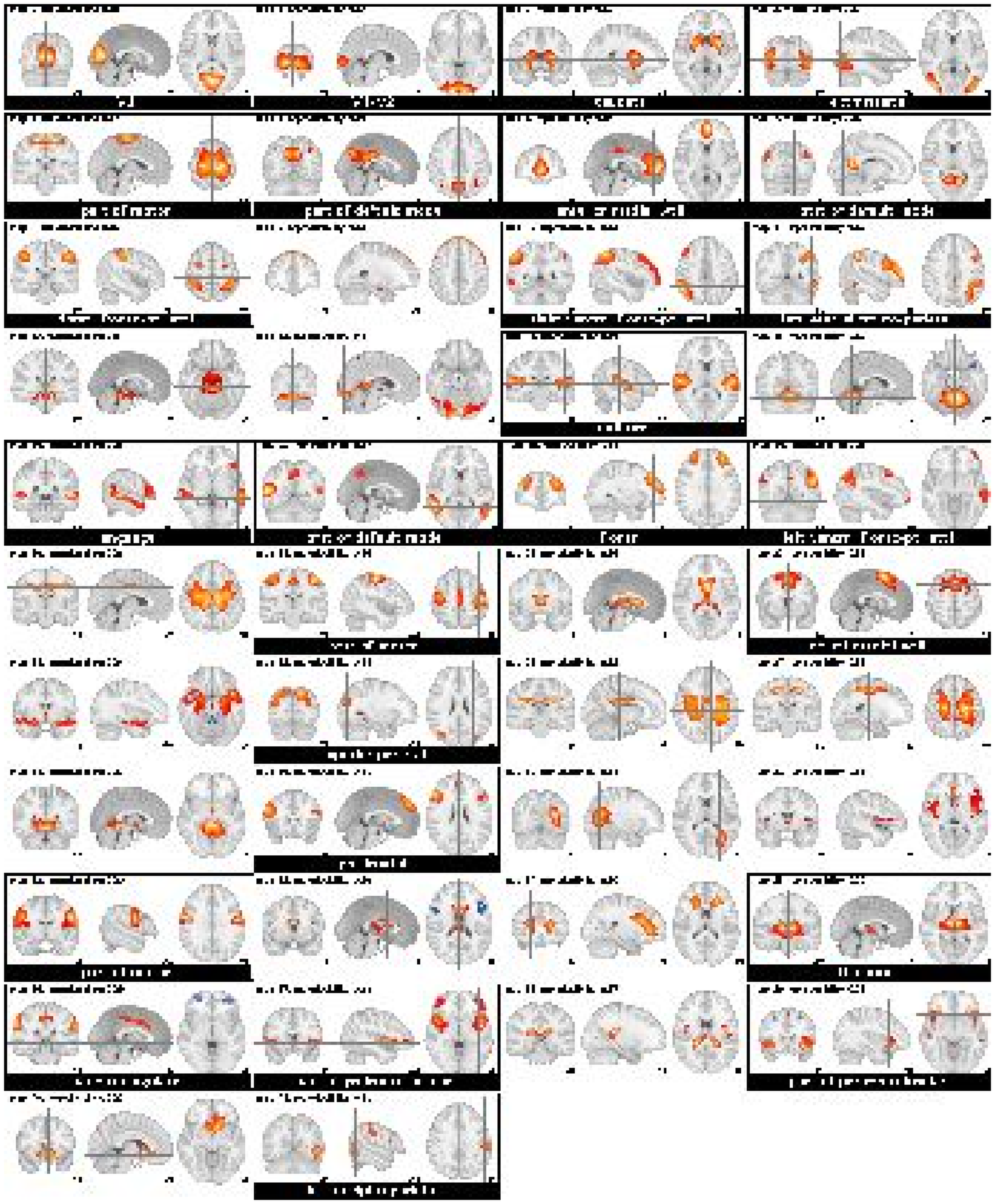}
   \caption{
	The 42 ICA maps extracted by CanICA from the resting-state dataset
(display in radiologic convention: the right hemisphere is on the left of
the axial view).
The maps are ordered by reproducibility (from left to right on each
line, and top to bottom line after line). Maps corresponding to 
functionally plausible networks are in a black frame, whereas maps 
likely corresponding to
artifacts are not framed. Extracted brain networks are labeled with the
name of the general structure they can be related to.
	\label{fig:all_ica_maps}
   }
  \end{center}
\end{figure*}

\begin{figure*}
  \begin{center}
   \includegraphics[width=\linewidth]{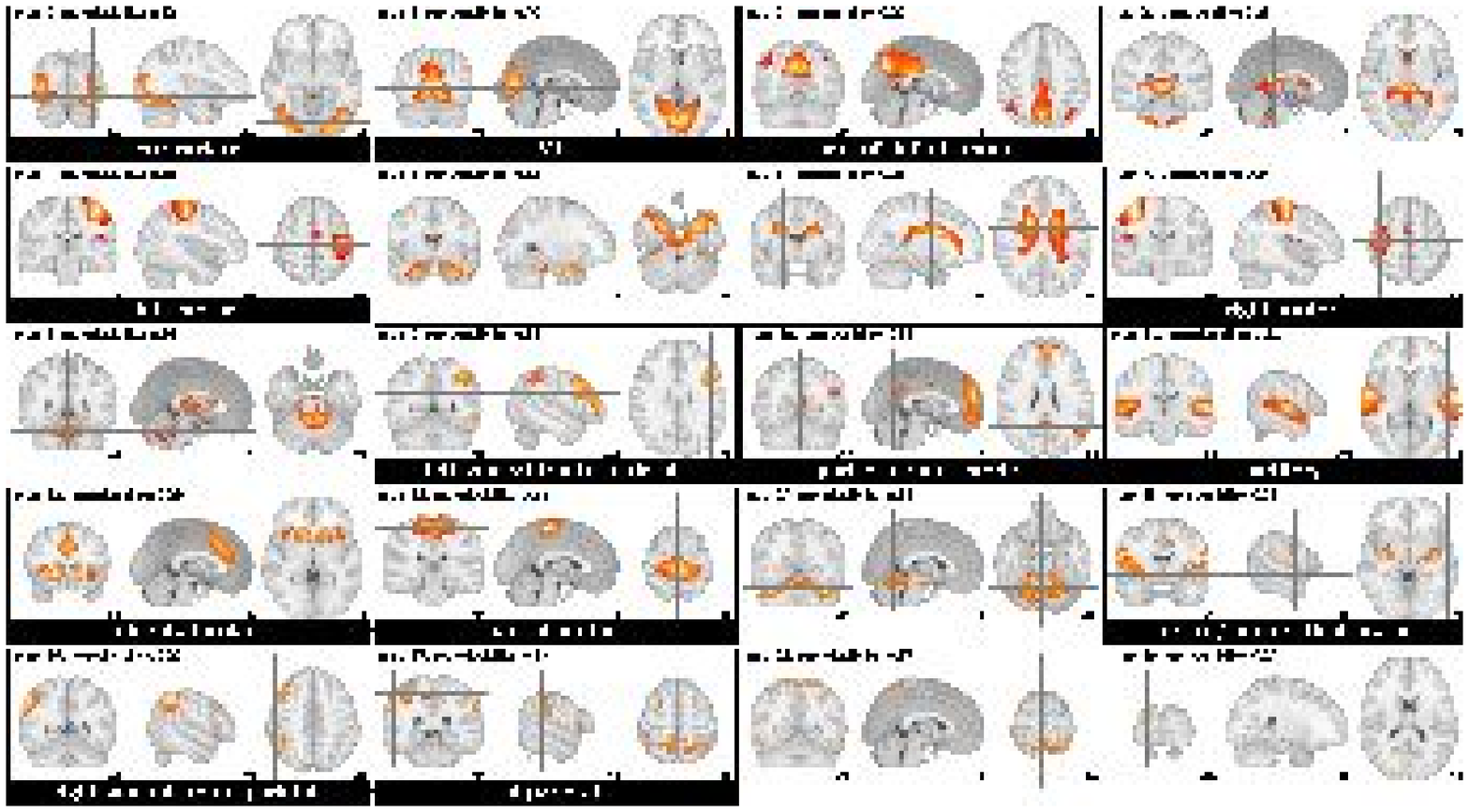}
   \caption{
	The 20 ICA maps extracted by CanICA on the functional localizer
dataset (radiologic convention).
	\label{fig:all_ica_maps_localizer}
   }
  \end{center}
\end{figure*}

\begin{table*}
   \begin{center}
   \end{center}
\def\mycol#1{& \bfseries\sffamily {#1}}
\def\s{&}

\begin{center}
\begin{tabular}{l|c|c|c|c|c}
	    \mycol{MELODIC   } \mycol{GIFT     } \mycol{MELODIC   } \mycol{CanICA  }   \mycol{CanICA } \\
	    \mycol{TensorICA } \mycol{GroupICA } \mycol{ConcatICA } \mycol{no CCA  }   \mycol{CCA    } \\
\hline      
	    	   \multicolumn{6}{c}{\bf Resting state \rule{0pt}{1.1em}}                             \\
\hline      
		   \multicolumn{6}{c}{Unthresholded ICA maps}				   	       \\
\hline      
$e$: subspace stability   \s  .47 (.06)    \s  .58 (.04)    \s  .58 (.04)       \s .36 (.02)     \s  .71 (.01)       \\
$t$: one-to-one matching  \s  .36 (.03)    \s  .53 (.04)    \s  .51 (.04)       \s .36 (.02)     \s  .72 (.05)       \\
\hline      
		    \multicolumn{6}{c}{Thresholded ICA maps}				   	       \\
\hline      
$t$: one-to-one matching  \s  .35 (.02)    \s  .10 (.01)     \s  .50 (.03)      \s .21 (.02)     \s  .62 (.04)       \\
\hline      
\hline      
	      	   \multicolumn{6}{c}{\bf Localizer \rule{0pt}{1.1em}}			   	       \\
\hline      
	      	   \multicolumn{6}{c}{Unthresholded ICA maps}					       \\
\hline      
$e$: subspace stability   \s .54 (.05)     \s     .25 (.03)  \s   .43 (.02)     \s .35 (.01)     \s  .52 (.01)       \\
$t$: one-to-one matching \s .36 (.02)     \s     .34 (.04)  \s   .35 (.03)     \s .37 (.02)     \s  .55 (.02)       \\
\hline      
	      	   \multicolumn{6}{c}{Thresholded ICA maps}			   		       \\
\hline      
$t$: one-to-one matching  \s .29 (.03)     \s     .02 (.01) \s    .31 (.05)     \s .26 (.03)     \s  .46 (.02)       \\
\end{tabular}
   \caption{
	Average reproducibility measures $e$ and $t$ for Group ICA, Tensor ICA
	and CanICA calculated on the half-split cross-correlation
	matrices, both for non-thresholded and thresholded maps. Numbers
	in parenthesis give the standard deviation across the different
	splits.
	\label{tab: measures}
   }
\end{center}
\end{table*}

\begin{table*}
\begin{center}
\small
\begin{tabular}{r|cc|cc|cc|cc}
			  & \multicolumn{2}{c|}{\sffamily\bfseries 12 subjects}   & \multicolumn{2}{c|}{\sffamily\bfseries 20 subjects} & \multicolumn{2}{c|}{\sffamily\bfseries 40 subjects} & \multicolumn{2}{c}{\sffamily\bfseries 62 subjects} \\
			  &  \sffamily\bfseries no CCA    & \sffamily\bfseries CCA       &  \sffamily\bfseries no CCA   &  \sffamily\bfseries   CCA  & \sffamily\bfseries  no CCA  \sffamily\bfseries  & \sffamily\bfseries CCA	& \sffamily\bfseries no CCA    & \sffamily\bfseries    CCA    \\
\hline			  
						\multicolumn{9}{c}{Unthresholded ICA maps}				   	  \\
\hline			  
$e$: subspace stability   &  .35 (.01) &  .52 (.01) &  .36 (.01) &  .57 (.01) &  .42 (.01) &  .71 (.01) &  .50 (.01) &  .78 (.01) \\
$t$: one-to-one matching  &  .37 (.02) &  .55 (.02) &  .40 (.02) &  .57 (.03) &  .45 (.01) &  .68 (.03) &  .49 (.01) &  .72 (.03) \\
\hline			  
						\multicolumn{9}{c}{Thresholded ICA maps}				   	  \\
\hline			  
$t$: one-to-one matching  &  .26 (.03) &  .46 (.02) &  .29 (.02) &  .45 (.03) &  .33 (.02) &  .56 (.03) &  .38 (.03) &  .60 (.04) \\
\end{tabular}
\caption{Reproducibility scores with using CanICA on the localizer dataset with many subjects.
\label{tab:many_subjects}}
\end{center}
\end{table*}

\paragraph{Resting-state dataset}

On the resting-state dataset, consisting of 820 scans, CanICA identified
in average 50 non-observation-noise principal components at the subject level
and a subspace of 42 reproducible
patterns at the group level (see equation \ref{eq:subject_variability}).
This number matches those commonly hand-selected by users in current ICA
studies. \citet{kiviniemi2009} reported 60 stable independent components
and 42 brain networks in a study of group-level ICA patterns stability on
a similar dataset. On this dataset made of long time sequences,
model-evidence-based methods such as those used in \citet{Beckmann2004}
select $200$ components at the subject level (or $340$ when
applied on smoothed data). Out of the 42 maps extracted by the CanICA, we
identified by eye 26 putatively components as brain networks (Fig.
\ref{fig:all_ica_maps}). When using CanICA without CCA, we putatively
identified only 11 components extracted as brain networks.
Other components related to physiological noise or movement form patterns
in the BOLD signal common to the group as they are related to
reproducible anatomical features.

\paragraph{Functional localizer dataset}

On the functional localizer dataset, consisting of 150 scans, CanICA
extracts 20 ICA patterns, out of which we identify 13 putative functional
networks (see Fig. \ref{fig:all_ica_maps_localizer}).  When using
equation \ref{eq:indiv_pattern_nocca} instead of equation
\ref{eq:indiv_patterns}, i.e. without the whitening of the individual
patterns imposed in the CCA, only 6 functional networks are identified
(see supplementary materials).

\subsection{Reproducibility results}

The reproducibility metrics we obtained by cross-validation are reported 
in Table \ref{tab: measures}. For the resting-state experiment,
on non-thresholded maps, the GIFT and ConcatICA perform similarly with
regards to subspace-stability, which can be explained by the fact that
they implement similar group models. 
For this experiment, the reproducibility of the TensorICA model is
not as good. Conversely, for the localizer experiment, the
reproducibility of TensorICA is good. The assumption of the TensorICA
model that networks share similar time courses across subjects is
clearly more suitable for task-driven studies. 
The CCA-based estimation procedure of the two-level group model of CanICA
yields the most stable subspace in both experiments. However, the
relative performance of different methods changes between the two
experiments that have very different signal length and TR. 
 
Performance of the method with regards to one-to-one reproducibility of
thresholded maps does not differ significantly from non-thresholded maps for
all the methods using histogram-based thresholding. 
On the other hand, GIFT thresholds a t-statistic map over back-reconstructed
subject-specific components, which is quite unstable across population
splits.
As a result, one-to-one matching of thresholded maps extracted by GIFT
does not perform well. We also note that the independent components
extracted from the functional localizer dataset are less contrasted
than those extracted from the resting-state data because the number of
volumes is smaller, and as a result thresholding has a more
detrimental impact on stability.

\begin{table}
\begin{center}
\small
\begin{tabular}{r|cc|cc|}
    & \multicolumn{2}{c}{\sffamily\bfseries Unthresholded} & 
      \multicolumn{2}{c|}{\sffamily\bfseries Thresholded} \\ 
			    &\sffamily\bfseries RS &\sffamily\bfseries Loc. 
	&\sffamily\bfseries RS  &\sffamily\bfseries Loc. \\
\hline
Matching above 50\% & 91\% & 64\% & 77\% & 44\% \\
Matching below 25\% & ~0\% & ~2\% & ~5\% & 25\% \\
Matching above 75\% & 55\% & 19\% & 34\% & 28\% \\
\end{tabular}
\end{center}
\caption{Percentiles of maximum one-to-one matching between maps
extracted with CanICA on two sub-group of 6 different subjects.
\label{tab:percentiles}}
\end{table}

The distribution of maximal component matching from one sub-group to
another provides an assessment of the reproducibility of the
individual maps. Table \ref{tab:percentiles} gives the percentile of
this distribution on both datasets, for thresholded and 
non-thresholded maps, when these are extracted with CanICA. The maps for
which there is a matching above 50\% are of particular interest, as
they have sufficient correspondence to establish a one-to-one
mapping with a simple matching scheme.

In addition, we have performed the same reproducibility analysis using a
higher number of independent components, to compare the different methods
for parameters similar to the analysis of \citet{kiviniemi2009} (see
table \ref{tab:many_ics} in the supplementary materials). As expected,
the reproducibility of the selected subspace is reduced compared to using
a small number of components selected, but the relative performance of
the methods is similar.

Finally, we have studied reproducibility for a large group of subjects,
on the localizer dataset for CanICA with and without CCA. As can be seen
on table \ref{tab:many_subjects}, reproducibility is improved when
estimating ICs on larger groups. On these groups, the use of CCA remains
an important factor of reproducibility.

We conclude from this cross-validation study that CanICA is a suitable
tool for purely data-driven extraction of stable markers from fMRI data
on a homogeneous group of subjects as it yields a small number of highly
repeatable features that can be identified between different groups. The
procedure is fully automatic as it does not rely on identification of
previously known activation patterns corresponding to cognitive paradigms. 
Unlike previous reproducibility
studies (e.g. \cite{Damoiseaux2006}), we report on the complete set of
patterns extracted.

\section{Discussion}
\label{sec:discussion}

\subsection{Factors impacting reproducibility}

\paragraph{Importance of the signal subspace}

We have used two different metrics to measure reproducibility of the ICs.
While $t$ measures exact one-to-one matching of the final ICs, $e$ is a
measure of reproducibility of the subspace, and is thus independent of
the ICA step. The fastICA algorithm was ran with different parameters (in
symmetric or in deflation mode, and using either the cube or the logcosh
non-linearity), and gave similar reproducibility results as measured by
$t$ on both datasets ($t$ metrics differing of .01). While GIFT and
MELODIC in TensorICA use different ICA algorithms, MELODIC in
concatenation mode and CanICA, with and without CCA, all use the fastICA
algorithm. We thus conjecture that the difference between the
reproducibility scores of these last methods are mainly due to different
subspace-selection procedures which consist of preprocessing and
group-model estimation.

\paragraph{Importance of thresholding on canonical correlations}

We compared 4 different sub-space selection procedures: CanICA with and
without CCA, MELODIC and GIFT. CanICA without
CCA implements a fixed-effect model, in which the principal components of
each subjects are concatenated without whitening before group-level
analysis. CanICA with CCA uses whitening of the individual datasets to
perform canonical correlation analysis and select group-level components
via a well-known reproducibility score: the canonical correlation. The
comparison of performing analysis with and without CCA over different
datasets (table \ref{tab: measures}), with varying group sizes (table
\ref{tab:many_subjects}), and for different number of ICs (table
\ref{tab:many_ics}, supplementary materials) shows the importance of the
CCA step. 

To our knowledge, there is no published detailed description of the
group-level estimation procedure implemented in MELODIC ConcatICA and
GIFT, we therefore base our discussion on our analysis of the software
packages. GIFT groups individual subject datasets and performs successive
PCAs and thresholding. This can be interpreted as applying nested
fixed-effects model. MELODIC applies a group-average filtering matrix
before performing whitening and group-level data reduction via SVD. For
these methods, the group-level components are thus not selected using
canonical correlation analysis on the individual subject data. While canonical
correlation selects the sub-space to optimize for reproducibility, the
procedures for group-level data reduction applied by both MELODIC and
GIFT impose some reproducibility criteria, as can be seen by their $e$
score on the resting-state dataset which out-performs a simple
concatenation (see table \ref{tab: measures}). 

We note that when performing analysis with a large number of ICs (table
\ref{tab:many_ics}, supplementary materials), selecting the signal
subspace on a canonical correlation criteria is less critical, as the
retained subspace covers a larger proportion of the initial signal. On
the opposite, on the localizer dataset, performing CCA is important for
reproducibility even for large groups of subjects (table
\ref{tab:many_subjects}).

\paragraph{Thresholding heuristic}

The metric $t$ applied to thresholded ICs measures the reproducibility of
features identified by the thresholding heuristic. This measure indicates
the ability of the ICA algorithm to yield reproducible salient features,
but is also a factor of the thresholding heuristic. The focus of this
paper is not to discuss thresholding heuristics, and we use a simple one.
However, we note that the impact of different heuristics (a mixture
model, as in MELODIC, or amplitude-based thresholding, as in CanICA) on
the cross-validation reproducibility score varies across the different
studies. On table \ref{tab: measures}, mixture models seem to impact
reproducibility to a lesser extend, whereas when using a high number of ICs (table
\ref{tab:many_ics}, supplementary materials), amplitude-based
thresholding performs better. This can be understood by the fact that the
heuristics depend strongly on the histogram of the maps, which vary with
the number of ICs or subjects. It is thus difficult to
conclude on thresholding without further study.

\subsection{Interpretation of the ICA patterns}

\paragraph{Sensitivity of the method to brain networks}

The patterns extracted display many different components that can be
interpreted as functional networks. On the resting-state data, this is
the case for 26 ICs out of 42.  
MELODIC's Concat-ICA, which corresponds to a state-of-the art
implementation of a one-level group model, when run on the same
dataset with the same model order, yields 20 identifiable functional
networks.
Using the same implementation as CanICA, with the same preprocessing and
thresholding steps, but without modeling two levels of variance 
(fixed-effect model), only
11 patterns extracted can be identified as brain networks.

However, the brain networks detected in both the fixed-effects and the
random-effects group models display close resemblance: modeling two
levels of variance, at the subject and at the group level, does not
significantly alter the group-level brain networks extracted. The
difference in extracted networks corresponds to a difference in
sensitivity of the methods to brain networks. This can be understood by
the fact that different networks can be activated in varying proportion in
each subject. This variability is tamed by the whitening introduced in
the CCA step that gives an equal weighting to all the information
captured in each dataset. As a result more networks are identified:
well-known networks can be segmented into reproducible sub-networks, and
additional networks seldom encountered in ICA analysis are extracted.

We note that on the resting-state dataset, because of the reduced field
of view, the upper part of the cortex is cropped (as can been seen on map
4), and we have no information on the networks in the lower temporal
lobes and in the cerebellum.

\paragraph{Extracted brain networks}
Most of the networks extracted can be related to stable networks
well-known from the resting-state literature, such as the visual system
or the fronto-parietal ventral and dorsal structures labeled as
attentional networks \citep{fox2006}. However, due to the increased
statistical power, we resolve sub-structures of these networks. 

For instance, the network known as \textit{default mode network} appears
to be split into several sub-networks separating portions of the
posterior cingulate region, as well as occipito-parietal junction,
precuneus and medial pre-frontal cortex (maps 5, 6, 7, and 18 on the
resting-state dataset and 2 and 10 on the localizer dataset). In
particular, the retrosplenial cortex stands separated from the posterior
cingulate cortex, grouped on map 7 of the resting-state dataset with
parietal regions. It has been shown to be the focus of anatomical
connections to the medial pre-frontal regions \citep{greicius2008}. Also
the ventral anterior cingulate cortex, shown on map 6 of the
resting-state dataset, has been consistently identified separately from
the default mode network, with evidence from EEG measurements that it
forms a differentiable network \citep{mantini2007}. This sub-network has
been associated with self-referential mental activity
\citep{dargembeau2005,johnson2006}. 

The networks identified in the frontal lobes, associated to executive
functions, display considerable variability in the literature as well as
between our datasets. Maps 37 and 39 on the resting-state dataset and
map 12 on the functional localizer correspond to the network related to
salience processing in \citet{seeley2007} and associated with task set
maintenance \citet{dosenbach2006,dosenbach2007,dosenbach2008}. As
reported by \citet{seeley2007,dosenbach2007}, it forms a distinct network
from the parietal-frontal network made of maps 10, 19 and 29 of the
resting-state dataset. 

Finally, we extract from the resting-state dataset some function-specific
networks such as the putative language network (map 16) or an
occipito-parietal network (map 41) related to the dorsal visual pathway.
Map 15 of the localizer dataset is a rich, anatomically-well-defined
cortico-subcortical motor network that comprises the ventral and cingular
motor cortices, as well as the insular cortex and the lentiform nucleus
and thalamus.


\subsection{Occurrence of the brain networks across groups}

Different networks can be recruited depending on the task, and thus we
may expect the occurrence of cognitive networks to vary when estimated on
different experiments, or different sessions.  
For instance, we observe that the right and left motor cortices appear on
different independent components for the localizer experiment, which may
be explained by the separate right and left finger-tapping tasks present
in the experimental paradigm of the localizer experiment, whereas in the
resting-state experiment, the motor areas are divided symmetrically in
somatotopic regions corresponding to lower body, upper body and face.

The reproducibility number, indicated on figures \ref{fig:all_ica_maps}
and \ref{fig:all_ica_maps_localizer}, gives an indication of the
occurrence of a network identified across subgroups. Visual and
attentional areas stand out among the most reproducible brain networks in
both datasets -- different parts of the visual cortex, and the network
coined {\sl visuo-spatial system} in \citet{beckmann2005}.
Parietal regions are embedded into various networks, bilateral or not,
and possibly but not necessarily associated with frontal regions,
according to the dataset considered or the processing strategy. The same
kind of observation also applies to frontal regions.
Bilateral auditory cortices, on the other hand, appear as a robust single
network across datasets.
In general, networks related to primary areas (sensory or motor) are
more stable across groups, paradigms, and methods, whereas networks
related to higher-level areas (mainly in the frontal lobes) or
higher-level specific tasks (such as language processing) are less
stable, and sometimes ill-resolved by the methods.

Although the comparison of resting-state versus activation datasets is
not the topic of this paper, one can notice that the results are
relatively consistent between the two datasets, and that the
resting-state data based on longer time series tends to produce a larger
number of interpretable components. This hints at the fact that these
networks may be defined not only at rest, but also in much more general
contexts of co-activation \citep{Smith2009}.

\subsection{Limitations of the method}

When running a group analysis using CanICA, signal originating from different
subject are put in correspondence via spatial alignment (performed by the
preprocessing steps). While the CanICA model accounts for subject-to-subject
differences, one of its major limitations is that it does not model
spatial variability across subjects. This is why the estimation is
applied on smoothed data. Similarly, the validation metrics $e$ and $t$ only
test for spatial correspondence. Differences between two ICs estimated on
different subgroups that can be accounted for by a spatial displacement of
features will induce poor reproducibility scores, and will hinder the matching
procedure used in the $t$ reproducibility score. 

A fundamental limitation of the ICA-based method presented and the
validation metrics is that they make no difference between neuronal and
artifactual signal. Thus, not only will the method extract and report
non-neuronal signal, but the corresponding maps will impact the
validation metrics. Additional pattern-matching techniques can be used
for these purposes \citep{demartino2007}. Extracting markers of the
non-neuronal signal can also have some value, for instance to use them as
nuisance regressors \citep{perlbarg2007}.

Choosing the model order of an ICA method is a difficult problem, as the
ICA algorithm will estimate orthogonal components spanning all the signal
subspace with no measure of statistical relevance. Our approach tries to
combine a criteria of non-Gaussianity at the subject level, and a
reproducibility threshold on the signal subspace at the group level to
identify the reproducible non-Gaussian signal subspace. One limitation of
selecting a subspace of reproducible signal across subjects is that
projection on this subspace may lead to projecting on the same IC
components that could be separated in larger subspace, as outlined by
\cite{kiviniemi2009}. This can be seen via component splitting when using
higher model orders. On the other hand, selecting a higher number of
components leads to less reproducible components and uncontrolled ICs
that are considered as artifacts and not interpreted in most analysis. 

In the CCA estimation step, the statistical measure of reproducibility
(canonical correlation) is a linear correlation measure and is not
informed with regards to the criteria of ICA, non-Gaussianity. This is a
limitation of the canonical correlation approach in our framework. One
could consider kernel or non-linear CCA models that would be more
specific to ICA criteria. Such procedures are much less
tractable, and there are less known results in statistics.

Finally, while CanICA models group-variability during the estimation step
to extract ICs representative of the group, it does not provide a
procedure to infer individual components related the group-level maps.
For this purpose, we suggest using dual regression as in
\cite{filippini2009}: the dual regression framework is independent of the
estimation procedure used to extract maps representative of the group.

\subsection{Algorithm complexity and numerical efficiency}

Because the estimation of the group-level model relies solely on simple
linear algebra routines, and the ICA optimization loop is performed on a
small number of selected components, it can be very efficient on large
data when implemented with optimized linear algebra packs, both in terms
of number of operations and memory. The computational cost of estimating
the full model on a group of $S$ subjects, with $m$ volumes each, and $p$
voxels in the brain is dominated by the cost of the subject-level PCA
that scales in $S \cdot m^2 \cdot p$. The group-level inference is made
of the CCA step, that scales in $S^2 \cdot p$, and the ICA step, that
scales in $S \cdot p$. For our data set, it takes a few minutes on a
$2\,\text{GHz}$ Intel core Duo. The speed of this step is critical for
cross-validation, as the initial subject-level model does not need to be
recomputed.

Performance is important to scale to long fMRI time series,
high-resolution data, or large groups. In addition, as the group-level
pattern extraction (CCA and ICA) is very fast, cross-validation of the
group patterns is feasible on modest hardware.

On the other hand, the model-order selection steps imply bootstrap analysis.
These steps are computationally costly. In particular estimation of the number
of principle components retained at the subject level involves evaluation of
many SVDs, a lengthy process (scaling in $p \cdot m^2$). We do not
estimate the subject-level model
order for each sub-group during the cross-validation of the group patterns.

\section{Conclusion}

We have presented a multivariate two-level generative model for
multi-subject datasets and applied it to an ICA model and corresponding
pattern-extraction algorithm for fMRI data, CanICA. Compared to existing
methods, our approach uses non-parametric noise description for
model-order selection. As a result, the method is auto-calibrated and
extracts in a fully-automated way meaningful and reproducible features
from fMRI data. In addition, we have introduced a cross-validation
procedure and associated metrics for ICA patterns and used it to
establish validity of group-level maps.

ICA is an unstable procedure with no intrinsic significance testing, but
we have shown that our pattern-extraction method, based on a
mixed-effects-like group model, can yield a set of thresholded maps of
which many are reproducible and can be identified one to one when
extracted from two different groups of only 6 healthy controls.
Reproducibility is an important feature of exploratory analysis methods,
as the validity of their results cannot be established by hypothesis
testing. In addition, group reproducibility on control groups and
one-to-one matching between groups is necessary when using extracted
patterns as bio-markers for group analysis.

\bibliographystyle{plainnat}
\footnotesize
\bibliography{restingstate}

\normalsize

\section{Supplementary materials}

\subsection{Group-level reproducibility threshold selection procedure}

Given individual datasets $\{\B{Y}_s, s=1\dots S\}$, after whitening 
using an SVD following equation \ref{subject_SVD}, for each subject $s$:
\begin{equation*}
    \B{Y}_s = \B{U}_s \, \B{\Sigma}_s \, \B{V}_s,
\end{equation*}
the first $n_\text{subj}$ components of $\B{V}_s$ for each subject for
the subject-level principal components $\{\B{P}_s, s=1\dots S\}$ 
and the remaining components form a basis of the observation noise
subspace, $(V_s)_{n_\text{subj}\dots m}$, where $m$ is the total number of
components, in other words the number of volumes acquired.

We wish to estimate the distribution of the maximum canonical correlation value
$z_\text{th}$ that is obtained if we select observation noise
rather than signal in $\B{P}_s$. For this, we use a resampling approach.
We generate datasets under the null hypothesis $\{\B{\tilde{P}}_s,
s=1\dots S\}$, such that $\B{\tilde{P}}_s\B{\tilde{P}}_s^T$, by drawing for each 
subject $n_\text{subj}$ components from
the subject's observation noise: $(V_s)_{n_\text{subj}\dots}$. 
On these datasets, we apply canonical correlation analysis
using an SVD:
\begin{equation}
    \B{\tilde{P}} = \B{\tilde{\Upsilon}} \, \B{\tilde{Z}} \, 
		\B{\tilde{\Theta}}.
\end{equation}
The maximum canonical correlation on the resampled dataset generated from
observation noise is thus given by $z_0 = \max\B{\tilde{Z}}$. Drawing many
realizations of the $\B{\tilde{P}}_s$ gives us access to the distribution
of $z_0$ when observing noise.

To control our null hypothesis to a p-value $p$, we take $z_\text{th}$ as 
the $1-p$ quantile of the distribution of $z_0$.

\begin{figure*}
  \begin{center}
   \includegraphics[width=\linewidth]{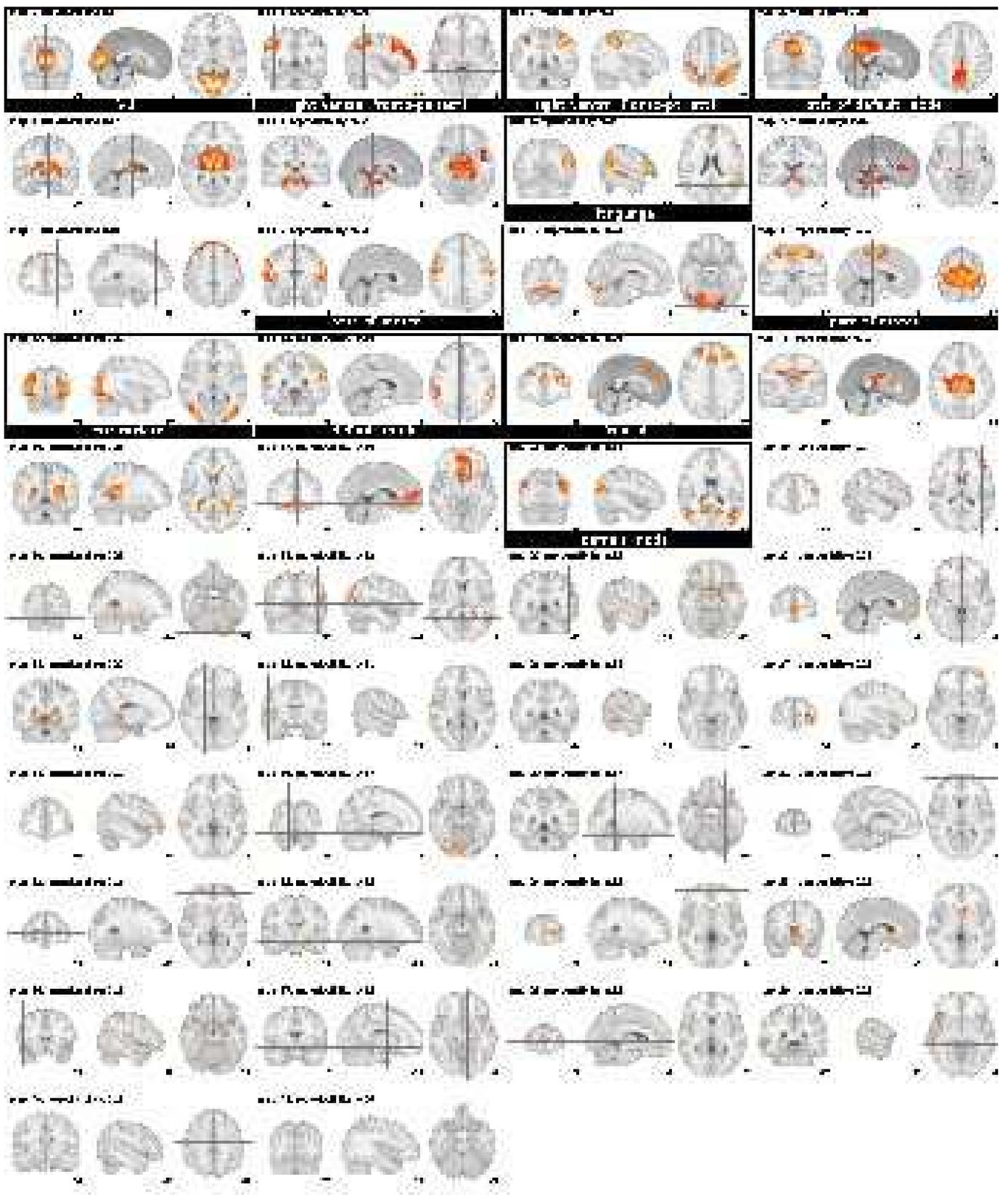}
   \caption{
	The 42 ICA maps extracted by CanICA on the resting-state dataset 
without CCA (radiologic convention).
\label{fig:all_ica_maps_no_cca}
   }
  \end{center}
\end{figure*}

\begin{figure*}
  \begin{center}
   \includegraphics[width=\linewidth]{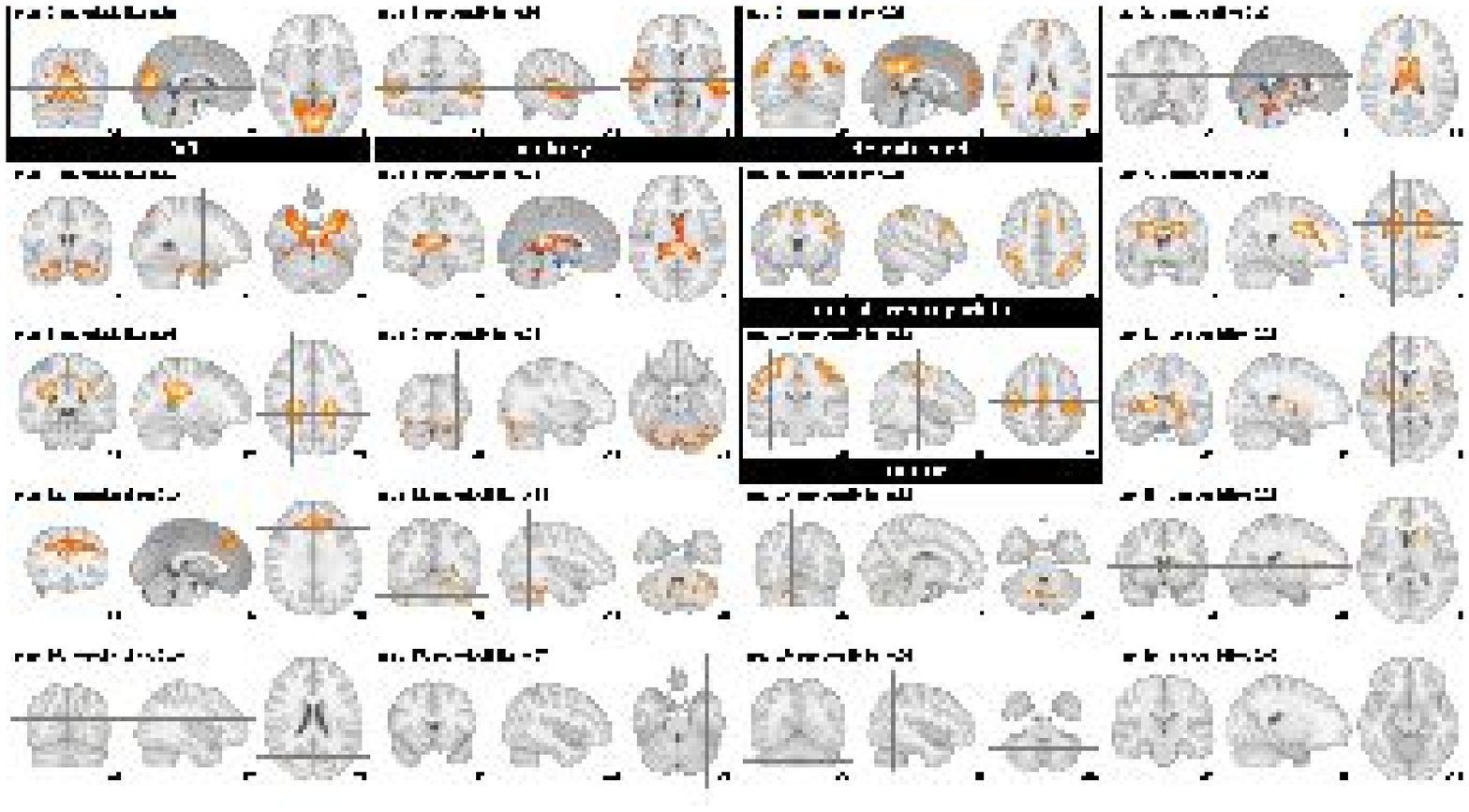}
   \caption{
	The 20 ICA maps extracted by CanICA on the functional localizer
dataset without CCA (radiologic convention).
	\label{fig:all_ica_maps_localizer_no_cca}
   }
  \end{center}
\end{figure*}

\begin{table*}
   \begin{center}
   \end{center}
\def\mycol#1{& \bfseries\sffamily {#1}}
\def\s{&}

\begin{center}
\begin{tabular}{l|c|c|c|c|c}
			\mycol{MELODIC   } \mycol{GIFT     } \mycol{MELODIC   } \mycol{CanICA  }   \mycol{CanICA }  \\
			\mycol{TensorICA } \mycol{GroupICA } \mycol{ConcatICA } \mycol{no CCA  }   \mycol{CCA    }  \\
\hline			
	    	   \multicolumn{6}{c}{\bf Resting state \rule{0pt}{1.1em}}					    \\
\hline			
		   \multicolumn{6}{c}{Unthresholded ICA maps}							    \\
\hline			
$e$: subspace stability   \s  .28 (.01)    \s  .41 (.01)    \s  .53 (.02)       \s .28 (.01)     \s  .49 (.02)      \\
$t$: one-to-one matching  \s  .15 (.01)    \s  .37 (.01)    \s  .47 (.02)       \s .28 (.01)     \s  .50 (.01)	    \\
\hline			
		    \multicolumn{6}{c}{Thresholded ICA maps}							    \\
\hline			
$t$: one-to-one matching  \s  .19 (.02)    \s  .06 (.01)     \s  .33 (.02)      \s .29 (.01)     \s  .57 (.02)      \\
\hline			
\hline			
	      	   \multicolumn{6}{c}{\bf Localizer \rule{0pt}{1.1em}}						    \\
\hline			
	      	   \multicolumn{6}{c}{Unthresholded ICA maps}							    \\
\hline			
$e$: subspace stability   \s .52 (.07)     \s     .17 (.02)  \s   .41 (.01)     \s .25 (.01)     \s  .30 (.01)      \\
$t$: one-to-one matching  \s .28 (.01)     \s     .21 (.02)  \s   .29 (.03)     \s .27 (.02)     \s  .34 (.02)      \\
\hline			
	      	   \multicolumn{6}{c}{Thresholded ICA maps}							    \\
\hline			
$t$: one-to-one matching  \s .23 (.02)     \s     .02 (.01) \s    .28 (.03)     \s .35 (.03)     \s  .39 (.01)      \\
\end{tabular}
   \caption{
	Average reproducibility measures $e$ and $t$ for Group ICA, Tensor ICA
	and CanICA calculated on the half-split cross-correlation
	matrices, with 70 ICs, both for non-thresholded and thresholded maps. Numbers
	in parenthesis give the standard deviation across the different
	splits. The localizer dataset is made of runs of 132 volumes,
	selecting 70 ICs explores thus the tail of the PCA.
	\label{tab:many_ics}
   }
\end{center}
\end{table*}

\end{document}